\documentclass[a4paper,12pt]{article}
\usepackage[utf8]{inputenc}

\title{String Theory for Metaphysicians}

\usepackage{url}
\usepackage{amsmath}
\usepackage{amsfonts}
\usepackage[font={small}]{caption}
\usepackage[margin=80pt]{geometry}
\usepackage{float}
\usepackage{braket}
\usepackage{natbib}
\usepackage{graphicx}
\usepackage[center]{titlesec}
\usepackage{blindtext}
\usepackage{hyperref}

\begin{document}
\author{Baptiste Le Bihan}

\date{2023}
\maketitle

%\tableofcontents{}

%\chapter{String Theory for Metaphysicians}

\begin{quote}
\textit{This is a draft chapter from a broader project on the metaphysics of quantum gravity. An earlier version was begun in 2021, circulated informally over the following years, and completed in its present form in 2023. A revised version will be incorporated into my forthcoming Cambridge Element, Metaphysics of String Theory. Please cite the forthcoming Elements rather than this draft, once available.
}  
\end{quote}

\begin{quote}
String theory is a minefield for philosophers in terms of conceptual difficulty and mathematical technicality. Presentations aimed at philosophers generally focus on providing precise and technical mathematical descriptions of toy models and inter-theoretical toy derivations to give an idea of what is at play in the field. This presentation takes the opposite approach, filtering out most mathematical considerations to provide an overview of the field, eyes on the ontology. The aim is to provide a compass for metaphysicians to find their way around string theory and the philosophy of string theory. The development of metaphysical arguments based on string theory against the fundamentality of spacetime is deferred to the next chapter.    
\end{quote}

%\abstract{String theory is a minefield for philosophers in terms of conceptual difficulty and mathematical technicality. Presentations aimed at philosophers generally focus on providing precise and technical mathematical descriptions of toy models and inter-theoretical toy derivations to give an idea of what is at play in the field. This presentation takes the opposite approach, filtering out the mathematical considerations to provide an overview of the field, eyes on the ontology. The aim is to provide a compass for metaphysicians to find their way around string theory and the philosophy of string theory. The development of metaphysical arguments based on string theory is deferred to the next chapter.}

\tableofcontents{}

\section{Introduction}

\subsection{In a Nutshell}
The purpose of this chapter is to introduce \textit{string theory}---the most popular extant research programme in quantum gravity---and some of its basic concepts in order to set the stage for the next chapter, which explains how the theory could imply the non-fundamentality of spacetime.\footnote{This chapter benefited greatly from discussions with physicist Emmanuel Fleurence who was kind enough to patiently answer many of my questions on matters difficult to find or understand in the literature, and from the feedback of James Read and the Geneva-Melbourne Metaphysics of Quantum Gravity discussion group. Many thanks to Sam Baron, Enrico Cinti, Jessica Pohlmann and Annica Vieser for their help.}$^,$\footnote{Good introductions into the subject are \citet{greene1999} and \citet{dummies}. For popular textbooks with introductions and passages that can be understood by philosophers not versed in string theory, see: \citet{green1987asuperstring, green1987bsuperstring,  polchinski1998string, polchinski1998superstring,  becker2006string, zwiebach2009first}. More recent textbooks are \citet{blumenhagen2013basic, tomasiello2022geometry}. See also on supergravity and M-theory: \citet{duff1999world}. Another important resource is \textit{nLab}, a collaborative wiki devoted to the mathematics of string theory, owned by Urs Schreiber and hosted by Carnegie Mellon University: \url{https://ncatlab.org/nlab/show/HomePage}. For a longer, and historical introduction to string theory from a philosophy of physics perspective, see \citet{rickles2014brief}.} As the subject is very rich, multifaceted and mathematically demanding, this overview has two main features. Firstly, it aims to give an idea of the big picture of the field, leaving out many details. Secondly, it is designed primarily for philosophers, and in particular metaphysicians, as an entry point into this difficult area. %It sets the stage for the next chapter. 
I hope that it will be taken by metaphysicians as an invitation to explore more widely the potential metaphysical implications of string theory, especially with respect to the nature of spacetime.

%Section \ref{HO} gives a brief overview of string theory. Section \ref{NP} sketches the naive ontology of string theory. Section \ref{Duality} complexifies the picture by introducing duality and its various ontological pictures. Section \ref{MA} has a more metaphysical flavour by discussing a number of arguments against the existence and fundamentality of spacetime.\footnote{My understanding of string theory follows mainly from \citet{witten1996reflections, greene1999, becker2006string, DH, DH2, HuggettDualities, Matsubara2017}
 %Becker, Greene, De Haro, Huggett, Rickles, Teh, Matsubara and Johansson, and Vistarini. 
 %The profane can find a good starting point in the introductory bits from \citet{becker2006string} and \citet{vistarini2019emergence}.} %\footnote{The beginning of part 2 and part 3 of this chapter are based on a joint work with James Read \citep{BLBJR} wherein we provide a classification of possible ontological interpretations of duality, rewritten and augmented with a discussion of the implication of each ontological interpretation for disputes in metaphysics regarding the nature of spacetime and composition. The resulting chapter differs greatly from the paper, and should not be taken to necessarily represent the current views of James Read.} 
%The relatively stable particles constituting matter around us would be in fact low-energy vibrations of those strings and branes. %High-energy vibrations are possible, but do not stay in existence long enough to be observed. 
According to the \textit{naive ontological picture}, or \textit{simple view}, presented by string theory reality is constituted by one-dimensional strings, and other higher-dimensional entities called `branes'. This view is naive because the ontology of string theory is much more complicated and difficult to grasp. However, it is useful from a pedagogical point of view to introduce this simple view first, and then to nuance and criticise it due to a number of complications. In the simple view, the elementary particles of the Standard Model of particle physics are regarded as underlying strings and branes in certain specific states of vibration, shape, length and other properties \citep[§21.4]{zwiebach2009first}. For example, strings appear more or less massive depending on their length, and the graviton---the possible carrier of the gravitational interaction---is identified with a closed string with low-energy vibration.

Moreover, reality is not made up of four spacetime dimensions (three spatial and one temporal) as, for reasons of mathematical consistency, it was necessary to introduce new dimensions resulting in mathematical spacetimes made of ten, eleven or even twelve dimensions. These different epistemic possibilities regarding the dimensionality of spacetime could seem puzzling at first sight. Doesn't string theory reveal the number of dimensions that structure the physical world, if the approach is correct? This is not the case, or at least not conclusively. %The reason why that's the case already anticipates the arguments in favour of the non-fundamentality of spacetime in string theory. 

The main issue with this, and with the simple view more generally, is that there is not one but five distinct string theories, which are based on different ontologies. These five string theories are all ten-dimensional but are conjectured to be limits of a more fundamental theory, \textit{M-theory}, which is eleven-dimensional (with ten spatial dimensions and one temporal dimension). One way to understand why M-theory is expected to be eleven-dimensional is that another limit of M-theory is eleven-dimensional \textit{supergravity}, a generalisation of general relativity that integrates supersymmetry (more on this below). To make the situation even more complex, one of the five string theories, \textit{type IIB superstring theory}, appears to also be a limit from a second, distinct, more fundamental theory: \textit{F-theory}, which has ten spatial dimensions and two temporal dimensions.\footnote{We will discuss in chapter \ref{fundamentality} how we should understand these relations of \textit{relative fundamentality} between theories. For now, it suffices to see that a theory $T_1$ is relatively more fundamental than a theory $T_2$ if, and only if, $T_2$ can be derived from $T_1$ using bridging principles between the primitive notions of the two theories, and mathematical procedures to ensure that the observational quantities predicted by $T_2$ approximate the one predicted by $T_1$ in relevant domains.} 

That the string world consists of strings and branes in a higher-dimensional spacetime is, however, an oversimplification. It replicates the common sense distinction between space and time on the one hand, and matter and objects on the other hand. But string theory does not fit this distinction very well---the situation is even worse than in general relativity, as we shall see. String theory embodies indeed an intriguing notion of great interest to philosophers and metaphysicians, the notion of \emph{duality}: a set of surprising equivalence relations between the solutions of theories previously thought to be different, and logically independent, candidate string theories. This makes it even more difficult than with general relativity to determine which part of the mathematics, if any, actually corresponds to a physical spacetime, and to identify the primitive strings.\footnote{For the reader schooled in string theory, I anticipate here various elements that will be introduced in the chapter. Firstly, T-duality, mirror symmetry and the AdS/CFT correspondence cast doubt on whether target space really represents spacetime. Secondly, the existence of strings is questionable: because of S-duality, which mixes string and brane states, in the context of the five superstring theories, and because M-branes (M2- and M5-branes) are the fundamental objects of M-theory, which no longer features strings.} More generally, string theory is nothing like a well-articulated physical theory from which we may directly read the ontology. Rather, string theory collectively refers to a set of mathematical fragments and results that, although highly suggestive of a few ideas, resist yet a full understanding in terms of a well-articulated ontology. Figure 1 shows some of the relations just mentioned, and the complexity of the big picture we have to deal with:

\begin{figure}[ht]
  \includegraphics[width=8cm]{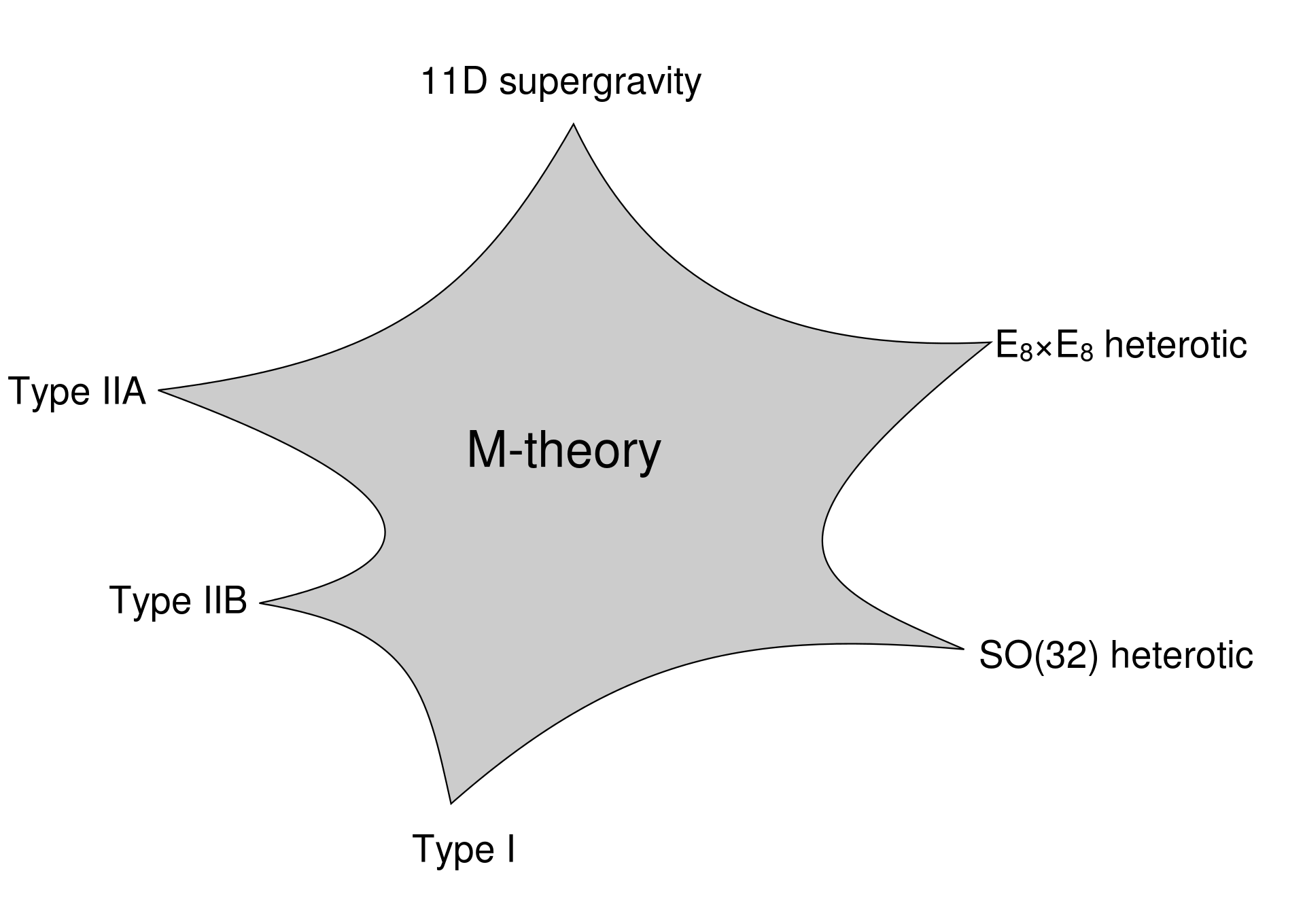}
  \centering
  \caption{M-theory, in the middle, is conjectured to be an eleven-dimensional theory. The five ten-dimensional string theories would be approximations of M-theory, and the eleven-dimensional supergravity would be another sort of classical approximation corresponding to the supersymmetric extension of general relativity. \\ \textit{Credit: Polytope24 \\ \url{https://commons.wikimedia.org/wiki/File:Limits_of_M-theory.png}}}
  \label{web}
\end{figure}

\subsection{History}

String theory arose from the attempt to explain the nature of the strong interaction and its structuring role in the existence of hadrons. These include mesons (pairs of quarks and antiquarks), baryons (triplets of quarks), and other exotic baryons created in laboratory like the tetraquarks and pentaquarks (respectively made of four and five quarks), all bound together by the strong interaction. In the late 1960s, it was postulated that the quarks making up the mesons were bound by strings, thus explaining the strong nuclear interaction, and why its strength does not diminish but rather augment with distance %\footnote{This historical presentation is based mainly on
\citep[p.~2-3]{becker2006string}.
%} %The central idea was that what looked like 0-dimensional particles was in fact 1-dimensional strings whose length is too small to be detected in most contexts. 
%Those strings were thought to have various properties. Like particles, they were supposed to move in space. But they also had intrinsic properties, namely topological and mechanical properties: strings could be either closed on themselves in a circle or open, and could oscillate with different frequencies. Depending on these properties, strings would appear in different forms, corresponding to the particles involved in the strong interaction---the hadrons. 
However, the candidate theory was eventually overtaken by another, more successful, theory of the strong interaction---\textit{quantum chromodynamics}---a quantum field theory, which is now an important part of the Standard Model of particle physics.%\footnote{The explanation for why the nuclear force between two particles does not diminish when the two particles get farther away is now that the more distance there}

String theory thus failed in its original form as a theory of the strong interaction.\footnote{Note, however, that quantum chromodynamics can be `recovered' from string theory via the AdS/CFT correspondence, to be introduced below. See, e.g., the holographic QCD entry on the nLab: \url{https://ncatlab.org/nlab/show/AdS-QCD+correspondence}.} But string theory proved to be a promising candidate for a more important endeavour: to serve as a theory of quantum gravity and perhaps even a final, ultimate, theory of the physical world. The central insight of string theory is that the particles and fields of the Standard Model of particle physics can be interpreted as vibrations of underlying, by many scales smaller, strings. Importantly, a certain mode of low-energy vibration of closed strings has the profile of a spin 2 massless particle, the expected profile of the graviton (but which interacts too weakly with matter to be detected by a physically reasonable detector). In other words, string theory seems to have the potential to explain the existence of the graviton, and thus serve as a quantum theory of gravity.

%Thus, the general idea that point particles could be grounded in one-dimensional entities remained. One reason for this is that one-dimensionality allows an answer to the problem of Lorentz-covariance introduced in the previous chapter*REF.

The ultimate hope within the string community is that string theory will provide a unification of the Standard Model of particle physics and general relativity, offering a framework capable of describing the whole of physical reality, from the smallest scale of description to the entire universe, thus providing an ultimate cosmological description of the physical world. Although much progress has been made in this direction, we are still far from a fully developed framework, and the theory remains yet to be empirically confirmed. General relativity can be derived to some extent from string theory but the Standard Model of particle physics has yet to be derived precisely from string theory (see, e.g., \citealt{huggett2015deriving} and\citealt{read2021landscape} for a philosophical discussion of the relation of string theory with respectively general relativity and quantum field theories). Overall, various key conceptual breakthroughs have taken place in the last decades, which are worth mentioning.

The first superstring revolution took place in the mid-1980s \citep{GREEN1984117, PhysRevLett.54.502, green1987asuperstring, green1987bsuperstring}, when it was realised that not one but five string theories could be developed.\footnote{I leave aside for the moment the bosonic theories of superstrings in order to simplify my presentation; I will discuss them in the next subsection.} Those are called \textit{superstring theories} as they integrate a novel form of symmetry, \textit{supersymmetry}, that posits the existence of superpartner particles for all bosons and fermions: each boson currently known should have a fermion partner, and each fermion currently known should have a boson partner.\footnote{A superstring theory capable of describing our world should not, however, be perfectly supersymmetric. The theory would have to account for spontaneous supersymmetry breaking, leading to the decay of superpartner particles at low energies; indeed, we do not observe any superpartner particles in the familiar low-energy world.} For instance, electrons should have `selectrons' as superpartners. 

As a matter of fact, since string theory continues the ontological lineage of quantum field theories, these particles should rather be considered as descriptions of underlying fields---not as primitive individuals.\footnote{I subscribe here to an ontological interpretation of quantum field theory in terms of fields, which I consider the standard and most natural understanding of quantum field theory. Not everyone agrees, however. For an opposing view, see \citet[section 5]{sep-quantum-field-theory} and references therein.} Yet, these fields must be distributed not in space, or in four-dimensional spacetime, but rather in a \textit{superspace}, namely a supersymmetric generalisation of a manifold, where the ordinary dimensions are associated to bosons, and other anticommuting dimensions correspond to fermions. That gives us a first example of how spacetime, as elaborated through general relativity, could be interpreted as non-fundamental: it could be that this superspace is the genuine arena of reality (see \citealt[pp.~65--68]{supersymmetry} for an introduction, and \citealt{MenonPBS} for philosophical analysis of the notion of superspace).\footnote{Supersymmetry can also be implemented in different spaces that we will encounter below, namely the target space and the worldsheet.} 

The five superstring theories are the \textit{type I}, \textit{IIA}, \textit{IIB}, \textit{heterotic SO(32)} and \textit{heterotic} $E_8\times E_8$ \textit{superstring theories}. They describe different objects with different properties; for example, type I, type IIA and IIB superstring theories include both closed and open strings, while the two heterotic theories contain only closed strings.\footnote{At least, this seems to be a standard view. See \citet{polchinski2006open} for an opposite view and discussions over NS5-branes in the heterotic theories.} %The five theories, for reasons of mathematical consistency, could not be four-dimensional, however. Further dimensions had to be introduced resulting in ten-dimensional theories, nine spatial dimensions and one temporal dimension.

The second superstring revolution brought novel ideas on the fore in the mid-90s. The central notion is \textit{duality} that we will discuss at length below, which roughly states that the five superstring theories, that were thought to be different alternative candidate theories to describe the world, turn out to be physically equivalent, namely to describe the same physical systems in different ways \citep{witten1995string}. The five superstring theories, and a sixth theory, eleven-dimensional \textit{supergravity}, were then taken to be related, in a not fully understood way, to a more fundamental theory, the eleven-dimensional \textit{M-theory}. Although M-theory is sometimes rather described as being on a par with the five superstring theories, it's fair to say that the majority view among string theorists is that M-theory should either be an \textit{absolutely fundamental theory} or, at least, \textit{relatively more fundamental than} the five superstring theories.

\subsection{Epistemological Status}

String theory remains the sociologically dominant approach to quantum gravity, but has been subject to sustained attacks. A number of them are worth mentioning.

A first well-known issue is the \textit{landscape problem}. The underlying idea of this so-called problem is the following. While physicists initially expected string theory to derive the Standard Model of particle physics in a unique way, the theory can actually be used to generate, depending on the geometric assumptions one uses (namely, on the selection of specific compactifications of dimensions into Calabi-Yau spaces), an incredibly high number of different quantum field theories. Calabi-Yau spaces are compactifications of six of the ten dimensions (they are closed on themselves with various shapes). The early hopes of string theorists were that it would be possible to find a few or unique ground state(s), because vacuum stabilisation would only work for particular cases of the wide range of possible geometries \citep{CANDELAS198546}. This hope has been dashed as analysis of vacuum stabilisation indicates the existence of a very large class of possible vacua (\citealt{KK, ashok2004counting, denef2004distributions}).\footnote{In addition to the fact that there are too many apparent solutions to describe our universe, there is the related question of the status of solutions that are clearly incompatible with the actual world (because they are not UV-complete, i.e, they are known to fail at describing our world at certain scales of energy): those solutions are referred to as the \textit{swampland}, and generate many questions about the relation between the landscape and the swampland.} 

However, this objection is not as powerful as one might think, as there doesn't appear to be anything specific to string theory in this respect. General relativity, for instance, has many solutions, and this does not seem to have bothered physicists in the past. %Think for example of the numerous solutions of general relativity. 
And if the difference between string theory and general relativity is claimed to be that string theory is supposed to be final, unlike general relativity, note the two questionable basic assumptions being made: first, that string theory will turn out to be a final theory and, second, that any final theory should not provide more than one solution, or at least not too many solutions. So the so-called landscape problem seems to be a false problem, standing on substantive metaphysical assumptions. What appears to be a real problem is to find at least one solution that is compatible with our current knowledge of the real world, which is simply the standard attempt to avoid falsification by collected data. For more details, see \citet{read2021landscape}; therein we discuss both the landscape problem and the additional claim that the landscape of solutions suggests the existence of a multiverse composed of many universes.

The second issue is the lack of predictivity of the theory, in the strong sense of offering novel predictions. This objection can be addressed in various ways. First, accommodations---or `retrodictions'---of data are also important confirmations of a theory, in a Bayesian sense. In fact, the epistemic nature of the distinction between prediction and accommodation is an important philosophical question (see for example \citealt{Hitchcock2004-HITPVA}), as it is not always clear why predictions are logically and epistemologically better than retrodictions. As far as retrodictions are concerned, string theory is incredibly successful since it is a quantum theory that replicates both the successes of general relativity in the relevant domain, and the successes of the Standard Model of particle physics by deriving some of its highly nontrivial general structures (such as the chiral fermions, the Higgs mechanism, or the hierarchy of couplings/masses), thus linking to all the empirical data in favour of them. Second, and related to this, it has been argued that string theory, in a Bayesian perspective, has a high likeliness of being true for reasons of consistency with everything we know about the actual world, a form of indirect, non-empirical, Bayesian confirmation (\citealt{Dawid2009-DAWOTC, dawid2013string, Dawid2013-DAWTAA-2, Dawid2019-DAWTSO-6}). Third, more neutrally, one could simply note that finding a specific empirical mark is a difficult challenge, not only for the string programme, but also for any theory of quantum gravity. The fact that the development of a phenomenology of string theory---the word `phenomenology' here refers to the field of physics whose aim is to relate physical theories to possible experiments and other empirical signatures---turns out to be difficult should not deter us from treating the programme seriously.\footnote{This list of objections is not intended to be objective. For other objections against string theory, see e.g. \citet{penrose2005road, penrose2016Fashion} and \citet{readpenrose} for a reconstruction and critical evaluation of some of the arguments provided by Penrose.}

%Perhaps more worrisome is the continued failure to detect supersymmetry at CERN. Although it is still possible to probe higher energies and find superpartner particles, this negative result certainly decreases the probability that string theory is true (however, the assessment of this probability decline is highly subjective, and differs widely from one physicist to another depending on their involvement in particular physics programmes).

String theory thus finds itself in an uncomfortable dialectical position: while remaining the sociologically dominant approach to quantum gravity, it is fair to say that it has suffered a severe blow in recent years, failing to find a single well-understood formalism or to formulate new testable empirical predictions. In my view, however, progress has been substantial conceptually, and the fact that the programme is proceeding slowly does not necessarily make it any less promising, especially considering the difficulty of the problems addressed, and bearing in mind that alternative approaches do not fare any better. In any case, and in keeping with my general agnostic attitude, I will not comment further on the epistemic status of string theory. We will now focus on the ontology associated with string theory and its possible implications for the non-fundamentality of spacetime.

%It also opened the possibility that the Calabi-Yau is actually larger than the spacetime in which we live. The idea behind \textit{brane cosmology} is thus that our ordinary space is in fact a 3-brane located within a space made of at least six dimensions. 

\section{The Simple View}{\label{SP}}

\subsection{Strings}

%According to string theory, the world is fundamentally made up of ten, eleven or twelve dimensions populated by one-dimensional and higher-dimensional entities vibrating at various frequencies, giving rise to what we consider to be the elementary particles of the Standard Model of particle physics. This picture is somewhat naive as the existence of \textit{duality relations} between apparently mutually inconsistent descriptions casts doubt on the ontologies of each of the theories considered separately. This section presents the naive picture by setting aside the role of duality, a notion that will be introduced in the next section. I will present in turn strings, branes, the different spaces where these entities live, two views about the relation of these entities, and conclude by explaining why quantum mechanics strongly suggests one of the two views.

This section presents the simple view by setting aside the role of duality, a notion that will be introduced in the next section, and the fact that we have more than one string theory. What I call the simple view can be interpreted as the application of a Quinean programme: aiming at reading directly the ontology from \textit{one} of the five theories, as in the form of its ontological commitments. For instance, what is the world like according to this specific string theory we consider?
All five string theories feature strings. Strings are one-dimensional entities, minuscule building blocks whose various properties give rise to various behaviours that are then conceptualised as a zero-dimensional point-like particles. Strings appear in the five superstring theories but %their role in M-theory is debatable. Some believe indeed that 
don't feature in M-theory; we will come back to this point below. %Strings being one-dimensional objects, 
%They are easier to quantise than the point-like particles, and 
Closed strings can have a mode of vibration that seems to correspond to the graviton. %particle potentially involved in the phenomenon of gravitation.
Strings have mainly three sorts of properties. They have a shape: they can be open or closed. When they are open, they have two end points that can move around more or less freely, or be trapped in some spacetime region. Second, they vibrate. Third, they interact by splitting and fusing with other strings in various ways. 

Strings are constructed in various ways depending on the theory considered. If at first, only the type I superstring theory seemed to include open strings, unlike the four other superstring theories, Polchinski and collaborators then showed that type IIA and IIB could in fact also include open strings, if their endings were attached to dynamical objects, the D-branes \citep{dai1989new, Polchinski1995}---we will return to this below. Now, the three theories are regarded as including both open and closed strings. The three theories obey different group of symmetries. The type IIA and type IIB theories differs in interesting ways. One is that type IIB string theory ascribes an intrinsic directionality to the strings (vibration patterns won't be the same in right and left directions, along a closed string). Another is that branes have an even number of (spatial) dimensions in type IIA and an odd one in type IIB.\footnote{If we factor in the extra temporal dimension, the classification gets reversed: type IIA branes have an odd number of dimensions, and type IIB branes an even number of dimensions.} To appreciate the characteristics of the two other superstring theories, the heterotic theories, we must mention another sort of string theories I haven't mentioned yet: the four\textit{ bosonic string theories}. These theories are 26-dimensional and do not describe fermions but only bosons, hence their name. For this reason, they do not constitute viable string theories to describe our world full of fermions. However, they are still considered useful and toy models for understanding some aspects and problems of string theories in general. 

The reason I mention bosonic string theories here is that they are used to construct two of the five superstring theories: the two \textit{heterotic} $E_8 \times E_8$ and \textit{SO}(32) theories. The term `heterotic' reflects that these theories are constructed by merging two different sorts of objects to generate a single theory. This is done by combining 26-dimensional bosonic strings with 10-dimensional supersymmetric superstrings.\footnote{``In addition to the type II theories, there are also two \textit{heterotic} superstring theories. These are remarkable closed string theories. While a type II closed superstring arises by combining together left-moving and right-moving copies of open superstrings, in the heterotic string we combine a left-moving open \textit{bosonic} string with a right-moving open \textit{superstring}! Out of the 26 left-moving bosonic coordinates of the bosonic factor only ten of them are matched by the right-moving bosonic coordinates of the superstring factor. As a result, this theory effectively lives in ten-dimensional spacetime. Heterotic strings come in two versions: $E_8 \times E_8$ type and \textit{SO}(32) type. These labels characterize the groups of symmetries that exist in the theories. $E_8$ is a group, in fact, it is the largest exceptional group (the \textit{E} is for exceptional). The group \textit{SO}(32) is the group generated by 32-by-32 matrices that are orthogonal and have unit determinant'' \citep[p.~324]{zwiebach2009first}.} Heterotic theories are hybrid creatures, and in this they will certainly not appeal to the metaphysician in search of a clean and simple ontology of string theory. For how can we trust such Frankenstein theories to accurately describe the real world? Perhaps this suggests that we should turn to `homothetic' theories (namely, type I, IIA and IIB) that do not use bosonic strings in their construction to seek metaphysical insight. Yet, these homothetic theories are rather convoluted and owe their salvation (i.e. consistency) only to the presence higher-dimensional entities, the branes to be introduced in the next subsection, which also signals a departure from the original hope of developing a theory with a relatively simple ontology.\footnote{Another issue raised by James Read (private discussion) is that the Standard Model of particle physics is likely to be recovered from a heterotic theory. These objections will lose some of their relevance when we move on to the complex view in the next section. It will then become clear that we should not try to decide which string theory is the most ontologically insightful, but rather view each of these theories as different windows onto a more fundamental framework.} %\footnote{Note that heterotic theories also include branes of a different kind. For the expert reader: I was implicitly referring to D-branes specifically when mentioning branes.} %, that could avoid the problem of UV divergence.

%Indeed, remember the material we covered on quantum field theory in section \ref{particlephysics}. Feynman diagrammes are used to catalogue all possible interactions between distinct particles, and between themselves.

%String theories are perturbative theories---they rely heavily on mathematical approximations. Just as quantum field theories, and the Feynman diagrammes, any interaction can be decomposed into a series of terms, each corresponding to a particular number of interactions. Indeed, as we saw in the previous chapter**REF, the so-called particles are in fact quanta of fields, and the continuity of the field generate an infinity of interactions, the particles interacting with themselves.

Unfortunately for the metaphysician, one can't simply assume the existence of strings, at least not at a fundamental level. Indeed, strings seem to make an appearance only in the five superstring theories that are perturbative string theories and not in the more fundamental non-perturbative theories (M-theory and F-theory) that feature higher-dimensional objects, as we will see below. That the five superstring theories are perturbative means basically that they rely heavily on mathematical approximations. Perturbation theory is a technique to approach equations that are too complex to be solved exactly. The technique is commonly used in quantum mechanics and quantum field theory. The idea is to first produce a first-order approximation of a solution to the equations, which is then corrected by another term, then another, and so on. These correcting terms in the series become less and less important (compared with the first terms in the series), their value not making much of a difference in the series overall, so that at a certain point the new terms are considered to be negligible. A conventional but non-arbitrary cut is thus applied to the series of terms to obtain an approximate final result. 

This might suggest that strings are not fundamental and only appear as an approximation of some deeper physics. %Logically speaking, it might still be the case that strings are fundamental, or only real at certain domains of energy, but their ontological status remains unclear. 
It's fair to say that since 1995 the dominating view in the string community is indeed that strings are not fundamental \citep{DuffMtheoryLayman}. Strings are now seen as perturbative approximations from the intersection of higher-dimensional, non-perturbative objects, the \textit{M-branes}. But `string theory' continues to be the term used to refer to the whole set of theories, whether or not they include strings. As there are different sorts of branes in string theory, let's start by introducing the branes involved in type I, IIA and IIB superstring theory: the \textit{D-branes}. We will encounter later the M-branes involved in M-theory.

%of  Thus, we should not be too quick in thinking that those strings, that correspond to interactions, are really building blocks of the world.

%\citep{witten1996reflections}.

\subsection{D-branes}

The `D-branes', `$p$-branes' or `D$p$-branes', are objects to which open strings can attach.\footnote{My presentation of the concept of branes closely follows \citet{vistarini2019emergence}. The reason for these names is that the two terms `D-branes' and `$p$-branes' were initially used to designate mathematical objects that were thought to be different, but that Polchinski shown to be identical \citep[pp.~194-194]{dummies}.} The central idea behind their introduction was that open strings might have their extremity attached to higher-dimensional structures subject to a dynamical evolution. The letter `D' stands for Dirichlet, referring to Dirichlet boundary conditions, a particular sort of mathematical constrains used to run calculations in differential geometry, and the letter $p$ for the number of spatial dimensions of the brane. A two-dimensional brane will thus be referred to as a D2-brane (a surface), and a one-dimensional brane as a D1-brane (or D1-string because of its dimensional similarity with strings). Sometimes they can be described as $p+1$ to clarify the number of spatial dimensions and emphasise the one temporal dimension. A particular string can have both ends attached to the same brane, or to different branes with possibly a different dimensionality. D-branes' (spatial) dimensionality ranges between 1 and 9, since the possible number of dimensions is capped by the dimensionality of the ten-dimensional (nine spatial dimensions) background space.\footnote{Other sorts of D-branes are D0-branes corresponding to points and D-1-branes, called instantons.} 
% that are not extended in space, but in a bundle space of the theory.
%One of those relations of duality, the AdS/CFT correspondence was made possible by the introduction of $D$-branes, a generalisation of the concept of strings to higher dimensions. 

\begin{figure}[ht]
  \includegraphics[width=8cm]{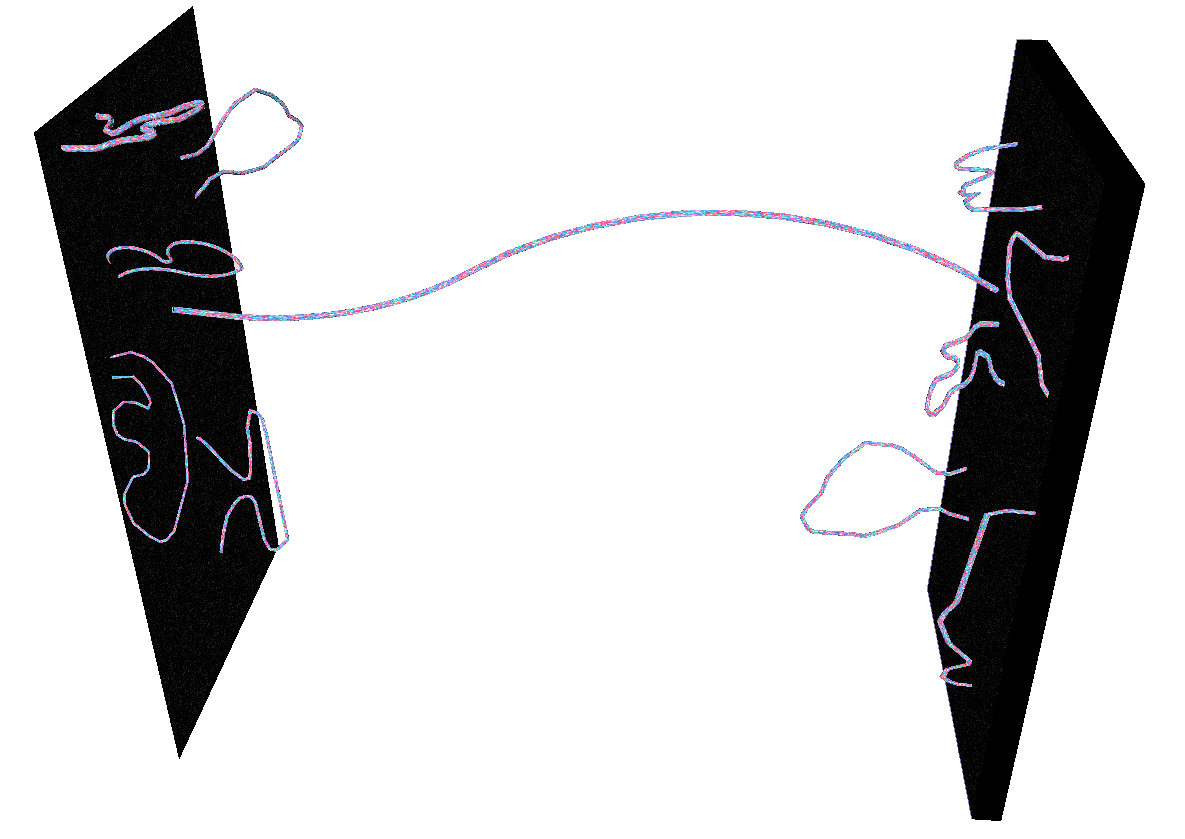}
  \centering
  \caption{A D2-brane (left) and a D3-brane (right) related by a string. Other strings have both ends trapped on the same D-brane. \textit{https://fr.wikipedia.org/wiki/D-brane}. }
  \label{D-brane}
\end{figure}

D-branes can carry various charges. These charges represent the possible conditions of string attachment between different branes. Indeed, the string endpoints also carry a charge, and the brane charges reflect the number of string endpoints they can accept, as a function of the value of the charge at the string endpoints. 

The specific types of charges borne by D-branes vary depending on the particular variant of string theory under consideration. These charges are associated with different fields and the topology of the branes, and include electrical and magnetic charges, as well as RR (Ramond-Ramond), NS (Neveu-Schwarz), topological, and flux charges. These charges play a crucial role in determining the possible configurations of string attachments between different branes. The endpoints of strings also possess a certain charge and the charges associated with D-branes indicate the number of string endpoints they can accommodate.

A second property of the D-branes is their tension. Tension is the property corresponding to the capacity of a D-brane to be influenced or deformed by an external object. If the tension is low, then a minor stress will have a significant effect on the D-brane. If it is high, on the contrary, then it will be difficult to affect or change the shape of the D-brane.

Another important property of D-branes is that they bind strings endpoints to certain geometrical shapes---they partially \textit{trap} them in regions of the background spacetime.  

Interestingly, this property of dynamic confinement has led to the \textit{brane cosmology}, according to which the extra dimensions of 10D space are not compactified with a small radius, but with a massive radius.\footnote{See, e.g., \citet{Langlois:2002bb}. Brane cosmology is also based on M-theory and its M-branes, to be introduced below.} Our three-dimensional space is then identified with a three-dimensional membrane floating in a higher-dimensional spacetime, called \textit{the bulk} or \textit{hyperspace}. Other three-dimensional spaces wander in the higher-dimensional spacetime, without any interaction except for gravitational interactions. In the words of Sen, it's...

\begin{quote}
a scenario in which the standard model fields live on a three brane, the directions transverse to the three brane being compact, and the directions tangential to the three brane describing the usual three dimensional space. There may also be other branes, separated from us in the extra directions, forming ‘shadow worlds’! \citep[p.~8]{sen1998recent}
\end{quote}

\noindent In this scenario, it becomes natural to interpret dark matter as the gravitational effect of other branes on our three-dimensional brane. There is then a simple explanation for the fact that we can feel the gravitational presence of the other branes even though we have no other kind of interaction with them and cannot see them. Recall that gravitation is described in this framework in the form of closed strings, and that closed strings have no extremities to tether to the branes. Thus, by their dynamic properties they are free beings that can navigate freely in hyperspace and interact with the different branes. Furthermore, brane cosmology provides a possible explanation for the hierarchy problem, namely why gravitation is so weak compared with the other fundamental forces. It is because it propagates in more dimensions over longer distances.

Let's set aside brane cosmology and focus on D-branes, more generally. The D-branes can be finite or infinite in size and appear in the three theories that involve open strings (type I, type IIA and IIB). Here is how Vistarini sums up the role of D-branes:

\begin{quote}
    D-branes are dynamical objects in their own right, and they are important in string theory for several reasons. The following is not an exhaustive list, but it can give a broad idea of why they are important. First, the ends of fundamental strings can attach to them. It has been showed that Yang-Mills type theories (like electromagnetism, weak and strong interactions) involve open strings that are attached to D-branes. Second, they carry the basic charges of a special class of fields which string theory necessarily describes. Finally, in a regime of strong string coupling, D-branes can become lighter than the string itself, and so their behavior can dominate some features of the low-energy physics. \citep[p.~60]{vistarini2019emergence}
\end{quote}

\noindent Branes are particularly important for homothetic theories (type I, IIA and IIB). In order to obtain a solution that resembles the Standard Model, the branes must be combined in various ways to create intersections, to match the profile of the Standard Model particles \citep[chapter 21]{zwiebach2009first}. Moving to metaphysical inquiry, one may then legitimate wonder about the ontological status of branes. Are those \textit{objects }in the same manner as strings, or merely locations in the background spacetime where string endpoints remain confined? Are they as fundamental as strings, or should we rather treat them as emerging from the configurations and properties of strings, in a way conceptually similar to how quasi-particles such as phonons are sometimes described as less fundamental than `real' particles?\footnote{For a philosophical presentation and discussion of phonons, see, e.g., \citet{Franklin2018-FRAEWL}.}  

This question is not settled yet as different school of thoughts cohabit on the fundamentality of branes. \textit{Brane democracy} refers to the view that strings and D-branes are on a par, in a sense to be made precise (see \citealt[p.~644]{taylor2004d}), a popular view.\footnote{``The different superstring theories each have different sets of (stable) D-branes, special branes that are defined by Dirichlet-type boundary conditions on strings. In particular, the IIA/IIB superstring theories contain (stable) D-branes of all even/odd dimensions. Each superstring theory also has a fundamental string and a Neveu-Schwarz five-brane. The branes of one theory can be related to the branes of another through the duality transformations mentioned above. Using an appropriate sequence of dualities, any brane can be mapped to any other brane, including the string itself. This suggests that none of these objects are really any more fundamental than any others; this idea is known as `brane democracy'.'' \citep[p.~644]{taylor2004d}.} However, \citet[section 2]{vistarini2017holographic} argues that the D-branes are not as fundamental as the strings, and that one should not conflate the \textit{epistemic democracy} of branes and strings with an \textit{ontic form of democracy}. Epistemic democracy refers to the fact that we need branes in string theory---in the sense that all string theories with open strings have automatically branes. Ontic democracy refers to the claim that strings and branes are equally fundamental, metaphysically speaking. Vistarini notes that it might well be that branes result from patterns of strings, generated dynamically by the influence of some laws of nature, resulting in the mere epistemic democracy of branes and strings, without any ontic democracy:

\begin{quote}
What exactly are D-branes in string theory? Are they fundamental in
the same way in which strings are? Are they also dynamical structures? Their ontological status in string theory is actually a matter of open debate. Two opposite views face each other. One assigning to them a fundamental status, similar to that enjoyed by strings, the other assigning a derived nature. In Vistarini (2017), I argue in favor of the thesis that D-branes are derived dynamical structures showing that they are not put in the theory by hand. \citep[88]{vistarini2019emergence}
\end{quote}

The claim that D-branes are not fundamental refers here to the view that D-branes are not part of the fundamental ontology of the superstring theories. D-branes would be non-fundamental, emergent structures generated dynamically by the behaviour of metaphysically fundamental strings. D-branes would hence be regions of variable dimensions of the background space, elected by the dynamically licensed motion of the strings' endpoints. Vistarini's argument is that in specific domains of energy, strings as a matter of fact are more fundamental than D-branes. She concludes that we should treat D-branes, in all circumstances, as being less fundamental than strings:

\begin{quote}
    What I mean is that from a perturbative point of view, that is, considering weak string coupling, as soon as we look at the D-branes’ quantum mechanical structure, strings are unambiguously more fundamental than them. Indeed, within this physical scenario, D-branes appear to be specific configurations of classical fields somehow emerging from quantum strings. \citep[88]{vistarini2019emergence}
\end{quote}

%\noindent Let's unpack this a bit. 

%The `weak string coupling' refers to weak interactions between the strings. Thus, an effective field theory perspective implies that the ontology is correctly described in a classical way at certain energy levels even though the description breaks down to higher energies. The view is that even though branes and strings might seem equally fundamental, branes are clearly not fundamental when interactions are weak. Thus,

%\begin{quote}
 %   Quantum mechanically they appear to be made out of composite string excitations becoming visible only at strong coupling regime. \citep[85]{vistarini2019emergence}
%\end{quote}

\noindent So, we would end up with a one-dimensional entity ontology, strings, which can be closed and open, and whose ends, when open, are confined to specific locations in the background spacetime. 

However, one might offer an argument in favour of brane democracy. D-branes seem to instantiate physical properties, and as such, one might claim that they are physical objects. If D-branes are mathematical constructs that represent the dynamical behaviour of physical strings, then how could they exemplify physical properties? %\footnote{My own view is that there are no physical objects at all according to our best physics, in the sense that what we call physical objects are physical fields that should be identified to collections of natural properties and natural relations, see, e.g., \citet{BLB2013, BLB2016philstud}.} 
The view is expressed for instance by \cite{dummies} who write:

\begin{quote}
Together, these two features of the D-branes---charge and tension---meant
that they aren’t just mathematical constructs, but are tangible objects in their
own right.
\end{quote}

However, it is questionable whether charge and tension are really physical properties typical of concrete physical objects. It may be thought that charge are associated to fields, and that subscribing to an ontology of fields uplift the pressure to regard those properties as signalling the existence of concrete physical object. It could be that, in the context of a field ontology, these properties do not need to be classified in the category of properties necessarily instantiated by objects. Perhaps D-branes, in the end, turn out to be regions of background space occupied by fields, and these fields conspire to produce the dynamic behaviour of D-branes, by trapping the extremities of strings. 

Setting aside the question of brane democracy, we end up with an ontology of strings, and possibly branes, living in a 10D or 11D spacetime. But already at the level of the simple view, that neglects duality and M-theory, a first complication arises. %This concludes my presentation of what I call the simple view of string theory, which neglects many substantive points that we are now going to explore.
One could view the strings, when extended in time, as generating two-dimensional worldsheets (by analogy with the one-dimensional wordlines in relativistic physics) that are more ontologically perspicuous than the background spacetime. By this I mean that the ontology of the world could be that of a worldsheet, such that the background spacetime would be a representational artefact of the mathematics we use to describe it.
%The first substantial modification to be made to the simple view relates to the `background spacetime'. 
As we shall see, the background spacetime might be a mere intellectual scaffolding useful for the construction of the theory, to be discarded when it comes to assessing the ontological implications of the theory. In fact, two elements call into question the reality of the background spacetime. One comes from the worldsheet perspective, relationist in spirit, the other from T-duality. Let's look at the first element in the next subsection and postpone duality for the next section.

%The second substantial change to the simple view is that strings do not really seem to be fundamental when we consider S-duality and the generalisation of the five perturbative string theories to non-perturbative frameworks, such as M-theory and F-theory. 
%Let's now unpack all of this.

\subsection{Target Space and the Worldsheet}

\begin{figure}[ht]
  \includegraphics[width=8cm]{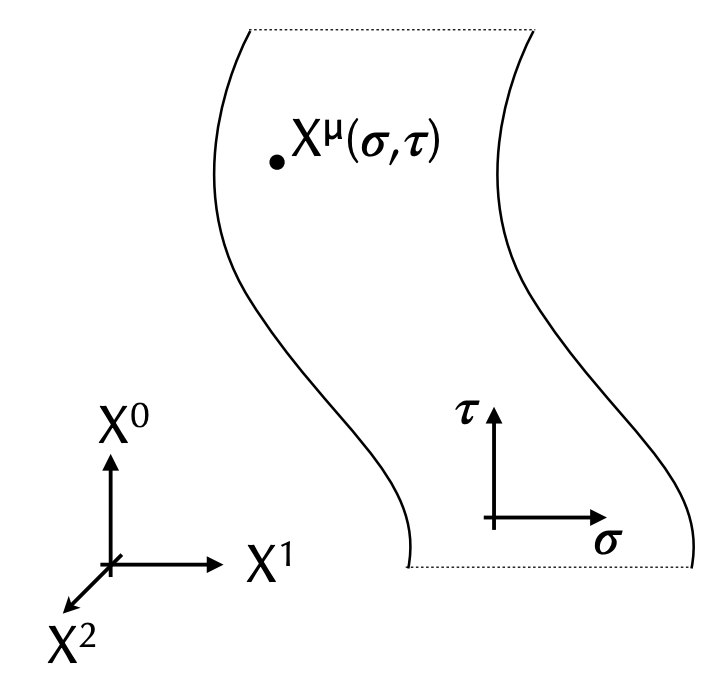}
  \centering
  \caption{A string, extends in the 10D background space, here represented as 3D (with dimensions $X^0$, $X^1$ and $X^2$) along the $X^0$ dimension (heuristically understood as an external time). It generates a 2D worldsheet. $\tau$ and $\sigma$ describe coordinates internal to the worldsheet, which can be heuristically assimilated to some internal time and internal (1D) space, respectively.\\
  \textit{Credit: \citet{huggett2015deriving} }}
  \label{worldsheet}
\end{figure}

%Perturbative superstring theories, rely on \alert{10D background spacetimes}. %Strings trajectories in the timelike directions of the embedding background spacetimes swept out a \textit{two-dimensional worldsheet}. 
The strings, by persisting in time (by extending along the time-like directions in 9+1 spacetime), generate a two-dimensional surface: the \textit{worldsheet}. In quantum field theory, particles were considered as point particles, whose trajectories in time generate wordlines. The metaphysician will find it natural to think of objects and particles as four-dimensional objects that persist in time by being spread over time. The resulting trajectories are lines, if the object in question has no spatial dimensions, and four-dimensional volumes if they have three spatial dimensions. In string theory, we have one-dimensional objects. Thus, when these objects persist in time (`time' refers to time-like directions in 9+1 spacetime, whose ontological status will be questioned below), they generate \textit{worldsheets}. If the string is open, then the worldsheet really does look like a sheet of paper (Figure \ref{worldsheet}). If the string is closed on itself, the worldsheet is more like a sheet of paper closed on itself, generating a \textit{worldtube} \citep{witten1996reflections, witten2015every}.

The events of string splitting and merging are now regarded as topological properties of the worldsheet. The resulting view is a complicated worldsheet, which can be represented as a complex network with pieces rolled up to form tubes. To keep things simple, let's continue to refer to this structure as a worldsheet.

Two metaphysical views are possible, regarding which of the background 10D spacetime and the 2D worldsheet is more fundamental than the other. One might expect that the background spacetime, although 10D, is quite `standard' by behaving somewhat like the Minkowski spacetime or special relativity or a pseudo-Riemannian spacetime of general relativity: it is simply a container in which things (understand strings) are located, just as in general relativity, one might naturally think of four-dimensional spacetime as the arena of reality. 

But of course, in general relativity already, the \textit{background independence principle} is supposed to show that things are not that easy, and that spacetime is not a \textit{fixed} background arena of the physical world in which things happen \citep{read2023}. If the general-relativistic spacetime were to be the arena of reality, then it would be a dynamic playground whose form fluctuates in concert with its material content. At best, then, the metric field (spacetime) is as fundamental as the matter fields (material entities), but certainly no more fundamental than the matter field. %are equally fundamental, might be more fundamental than the geometrical arena (described by the manifold and the metric field). %But, at least, this is the only candidate for being the arena of reality.\footnote{If we set aside exotic approaches that identify the arena of reality to configuration of other modal spaces, under the influence of the wave function realist approach to quantum mechanics.} 
However, in general relativity, the metric field remains a genuine source of influence on matter---perhaps even causal---and thus appears to have a reality of its own. 

In string theory, the existential crisis of spacetime is similar but more radical to some degree as the 10D spacetime in which strings are supposedly `embedded', appears to dynamically emerge, in a relationalist spirit, from quantitative properties located on the two-dimensional worldsheet
%---opening the way for the view that the background spacetime simply does not exist 
(see \citealt{witten2015every} for a defence of this point of view in physics, and e.g. \citealt{huggett2015deriving, baker2016, vistarini2019emergence, READ2019103} for philosophical discussion).\footnote{My presentation might suggest that reasons to doubt the fundamentality of the target space in string theory are identical to the reasons to doubt the fundamentality of spacetime in general relativity (namely to doubt metric substantivalism). But, it's not the case as we shall see. For a longer discussion of background independence in physics in general, and in the context of string theory, see, e.g., \citet{read2023}.}
What's really fundamental in this picture, are the strings (not the branes), which when considered in their historical totality, including their interactions, form a universal worldsheet (or quantum superposition of worldsheets). The worldsheet approach thus views the world as a 2-dimensional reality, on which quantum fields fluctuate (or as quantum superposition of 2-dimensional states).\footnote{The problem of measurement of quantum mechanics (see chapter ***REF) appears again here, and depending on the approach one takes to the ontological interpretation of quantum mechanics, one will also end up with different ontologies in the context of string theory. It may be that string theory will one day offer us reasons to favour one approach to the ontology of quantum mechanics over another, but in the present state of our knowledge, it may also, on the contrary, be the case that string theory needs to be complemented by a particular solution to the measurement problem.} Indeed, all the information contained in the background spacetime and the locations of strings in it, are translated in information about scalar fields (mathematical descriptions that can naturally be physically interpreted as distributions of quantitative physical properties) located on the worldsheet.\footnote{Mathematically, scalar fields are functions from points in space to real numbers.} %This sort of reasoning is an instance of \textit{dimensional reduction}, which is one mathematical recipe to move from a description involving a specific number of spatial dimensions to a description involving a different number of spatial dimensions---another mathematical recipe used to change the number of spatial dimensions in a mathematical description, is \textit{compactification}, that we will encounter below.

In arguing that the worldsheet approach is relationist in inspiration, I am referring to the relationist thesis, which has been discussed at length in the classical debate about whether space and time are absolute substances in which material objects are located, or collections of relations between material objects, especially in the context of Newtonian mechanics and, later, general relativity. Relationism can be rigorously defined in various ways. I follow here my preferred definition, assuming that relationism has one main defining feature, which implies two other features \citep{BLB2016philstud}. The primary feature is that space and time are collections of relations, not substances. This in turn implies, first, that these relations depend ontologically on the existence of their relata, namely material entities (usually taken to be material `objects', but as I argued in \citealt{BLB2016philstud}, there are strong reasons to think that relata should rather be classified as properties, not objects). This also implies that space and time are not `containers' (a perhaps misleading but common metaphor) in which events occur and, more importantly, that space and time are constructed from more fundamental ingredients, namely the collection of material entities (as described as matter fields in general relativity).

Similarly, in string theory, strings could be more fundamental than the background spacetime on which they live. This spacetime, in the string jargon, is called `target space'. %How we call them does not matter of course, but we will need to keep in mind different notions of space, so I will continue to call it target space to differentiate it from other notions of space. 
Naively, the target space is just the real physical space, or taking into account the background independence in general relativity, the natural mathematical description in terms of a manifold and a metric field, of the structure we measure with rods and clocks (the operational or phenomenological spacetime).\footnote{I'm using the expression `phenomenological space/spacetime' in the sense of \citet[p.~83]{HuggettDualities}. Note that the word `phenomenology' describes two very different fields in physics and philosophy. In physics, it refers to the field aimed at building bridges between theoretical principles and possible empirical verifications. In philosophy, the term refers to the analysis of conscious experiences as they appear to us, i.e. as phenomena in contrast to the noumenal world or world in itself, in line with Kant's famous distinction in the \textit{Critique of Pure Reason}, in 1781.} As we will see in the next section and the next chapter, there are reasons to doubt that the mathematical target space describes directly, in a faithful way, the physical space (the arena of reality). The view makes it very hard to clearly separate between spacetime and matter, insofar as the worldsheet in some sense is both material and `dimensional' (spatiotemporal in some limited sense).\footnote{See \citet{BLBEJPS} for more detail on the worldsheet approach and how it relates to relationism and the non-fundamentality of spacetime.}

%However, we will see that duality anyway sheds doubt on the naive 10D picture. %This is because different 10D pictures turn out to be equivalent (because of T-duality and mirror symmetry), suggesting that the 10D spacetime are pure mathematical artefacts. Likewise, various 2D pictures appears.
Is the worldsheet approach more than simply a theoretical curiosity? We will return to this question in the next chapter. But let's note already that a number of physicists, and especially its champion Edward Witten, have at least flirted with the view that the worldsheet approach is a better interpretation of the physics at hand than the spacetime view. Here is how he puts the point:

\begin{quote}
    [I]n string theory one does not really have a classical spacetime, but only the corresponding two-dimensional field theory; two apparently different spacetimes $W$ and $Y$ might correspond to equivalent two-dimensional field theories. \citep[p.~29]{witten1996reflections}
\end{quote}

\noindent If different background spacetimes correspond to the same physics, then surely the background spacetime should not be seen as accurately describing the fundamental nature of reality---for that, we should probably turn to the ontology of the worldsheet. It also seems that the worldsheet approach can describe tensionless strings, when the standard `spacetime'  approach cannot. This might suggest that the former is relatively more fundamental than the second, in the sense of getting closer to the truth.\footnote{Thanks to Enrico Cinti for bringing this point to my attention.}   In fact, as we will see below, the fate of spacetime, would it be in the guise of the target space, or as the phenomenological spacetime we measure via rods and cloaks, is deeply tied to a resolution of the ontological significance of those relations between equivalent descriptions that rely on different spacetimes. Those strange, seemingly metaphysically-loaded, equivalence relations belong to a broader family of relations called `duality', that we will shortly introduce in a semi-technical way. For now, we set aside the worldsheet approach as a possible way to read the ontology of a superstring theory.

%Quantum mechanics is well-known to posit a duality between position and momentum: the more precise a measurement of momentum of a particle is, the less precise our knowledge of its position will be, and vice versa. The thing is, in quantum string theory, this duality generalise and imply that it will always be equivalent to measure momentum and position in different spaces (\citealt{witten1996reflections}; see \citealt{HuggettDualities} for discussion). Now, the sketch of T-duality I have just offered rely on a mainstream, but not universally shared, understanding of duality. In the next section, we will introduce more precisely the notion of duality, present the most famous examples (T-duality, mirror symmetry, S-duality and the AdS/CFT correspondence), and present a taxonomy of possible ontological interpretations.

%What about the 2D picture? It's difficult to understand how exactly the 2D picture is affected by T-duality and S-duality. It might well be that the 2D picture can resist the ontological devastation prompted by duality relations, but it remains to be seen if that's really the case.

\section{The Complex View}{\label{Duality}}

What I call the \textit{complex view of string theory} moves from evaluating the ontological commitments of a specific string theory (what the world is like, ontologically speaking, according to the theory considered) to evaluating those of all theories, considered as a whole. The aim is to understand the significance of the existence of a plurality of distinct theories, and of the fact that each of these theories is linked to some of the others by interesting non-trivial relations. As we mentioned before, those duality relations might suggest the the different theories describe some different aspects, and some similar aspects but in different ways, of a relatively more fundamental theory, the so-called\textit{ M-theory}. This section introduces provides a description of the four main kinds of dualities before turning to M-theory and F-theory.

\subsection{What's a Duality?}

In order to introduce the notion of duality between theories, it is useful to say a few words about the nature of theories. Many views on the topic coexist in the literature. Some are logically well-articulated views, such as the \textit{semantic} and \textit{axiomatic} views which have been at the centre of discussions on the nature of scientific theories \citep{fraassen1987semantic, van2014one, halvorson2012scientific, halvorson2013semantic, Lutz2017-LUTWWT}. Others lean more towards the empirical side and see theories via their functional role in various contexts \citep{functionaltheory}, or  as classifications used scientists to unify their body of knowledge about a phenomenon \citep{duhem1906theorie}. In this tradition, Steven French has recently argued that, strictly speaking, theories do not exist \citep{french2020there}. In what follows, I will give a definition of duality by using a semantic approach to theories. This choice doesn't reflect any principled reason; it's simply that the semantic approach drastically eases the introduction and discussion of dualities.

According to the semantic view, a theory can be identified to a collection of `models', which in this context, refer to solutions of some equations. To quote van Fraassen, a model is ``Any structure which satisfies the axioms of a theory'' \cite[p.~43]{BVF}. These models are generally structured in a model space as follows. A set of dynamic models is a subset of the set of kinematic models. Kinematic models obey a first class of constraints, while dynamic models correspond to the solutions of the theory's dynamic laws, a second layer of constraints.
%Whether those solutions are three-dimensional or four-dimensional to include a complete story of the described system leads to important subtleties.
%This picture in mind, one can then enquire about which of the two is more fundamental. Is the theory more fundamental than the set of its models? Or is it less fundamental than the theory, either because it's just identical to it, or because it's eliminated in favour of the class of models?
%Associate with every theory a class of `models', equivalently `solutions'.  (Roughly speaking, physical theories are specified by certain dynamical equations; each model represents one possible solution to those equations---hence the latter choice of nomenclature.)
We can then define the notion of empirical equivalence. Two solutions are `empirically equivalent' if and only if they agree on all `physically observable data', namely on empirical substructures, to use the expression of \citet[p.~64]{BVF}.

A duality is a mapping---a systematic correspondence---between the spaces of solutions of two theories, such that models related by that map are physically equivalent.\footnote{See, e.g.,
%\label{fn-gauge}Such is the definition of dualities presented in e.g.
~\cite{Matsubara, Read}. %, which will suffice for our purposes. %Note that one might augment the criterion of `empirical equivalence' by requiring that all quantities regarded as being physically meaningful (whether observable or not) be preserved under the duality map. %We agree that the most striking examples of dualities---including examples of dualities from string theory---satisfy this stronger condition. However, in this paper we choose to work with the weaker notion of a duality proceeding in terms of empirical equivalence alone, for this will suffice to make all necessary points regarding the interpretation and ontology of duality-related theories. 
%Note also that this is the way philosophers talk about duality---physicists are more likely to say that we have just one theory, described in different ways. 
For more detailed and comprehensive approaches to the definition of dualities, see e.g.~\cite{DH, DH2}. Note also that it need not be the case that the duality map is one-one---it might instead be that a class of solutions of the first theory is mapped to a single solution of the second theory under the duality. %This will be of relevance below.
}
The condition of physical equivalence can be broken down into two parts: \textit{empirical equivalence} and \textit{theoretical-quantities equivalence}. Empirical equivalence states all observable quantities of the two theories are the same for any pair of duality-related models of the two theories. Physical equivalence is stronger in adding that the non-observable theoretical quantities (that are standardly regarded as physically meaningful) are also the same (at least in the sense of having the same values). Thus, physical equivalence goes one step further than empirical equivalence in making the equivalence more ubiquitous and structural, with a systematic mapping of all the physically meaningful quantities of the two theories.

Dualities should also be contrasted to the standard notion of symmetries, even though, in the end, dualities can be regarded as certain sort of symmetries, as long as we distinguish between \textit{intra}-theoretic symmetries and \textit{inter}-theoretic dualities.\footnote{For a more systematic discussion of the relations between symmetry and duality, see \citet{DeHaroButterfieldSymDua}.} As \citeauthor{DeHaroButterfieldSymDua} write,

\begin{quote}
in the literature, it is agreed by all hands that a duality is like a `giant symmetry': a symmetry between theories. For in physics, the basic idea of a symmetry is a map taking a state of the system into another appropriately related state; and correspondingly mapping physical quantities [\dots]. And in a duality, an entire theory is mapped into another appropriately related theory. \citep[p.~2974]{DeHaroButterfieldSymDua}
\end{quote}

\noindent Symmetries qualify possible transformations between equivalent states inside the solution space of a given theory. Dualities apply to possible transformations between the solution space of distinct theories. (An important qualification here is that dualities in fact can relate a theory to itself. That's the case of type IIB string theory, which is self-dual, via S-duality.)

The four main dualities in string theory are: T-duality, mirror symmetry, the `AdS/CFT correspondence and S-duality. Let's review them in turn. %Other dualities are sometimes discussed that I mention at the end of this section.

\subsection{T-duality and Mirror Symmetry}
   
\textit{T-duality} is the abbreviation for \textit{target-space duality}, and thus relates to the target spaces (the background spacetimes wherein the strings live) involved in different string theories.
%Let's give a first, intuitive sense, of T-duality, as it's our duality along the road. 
A model of a theory in which the strings wind up within a dimension whose radius of compactification measures $R$, corresponds by T-duality to a model of the dual theory in which the strings are wound within a dimension whose radius of compactification measures %$l_{s}^2/R$ 
$1/R$. 
%(where $l_s$ is the string length). 
Whether the dimensions are compactified in such a way that the strings closed in on themselves describe a radius of size $R$, or on the contrary a radius of size 
$1/R$,
%$l_{s}^2/R$, 
the world described is exactly the same \citep[p.~481]{Matsubara}. Put another way, a world in which the strings are free (i.e. don't wind up) within a dimension D having a radius R is perfectly equivalent to a world in which the strings wind up within a dimension D' having a $1/R$ compactification radius.

Consider the following useful visual representation (Figure \ref{T-duality}), and description both offered by Nick Huggett:

\begin{figure}[ht]
  \includegraphics[width=8cm]{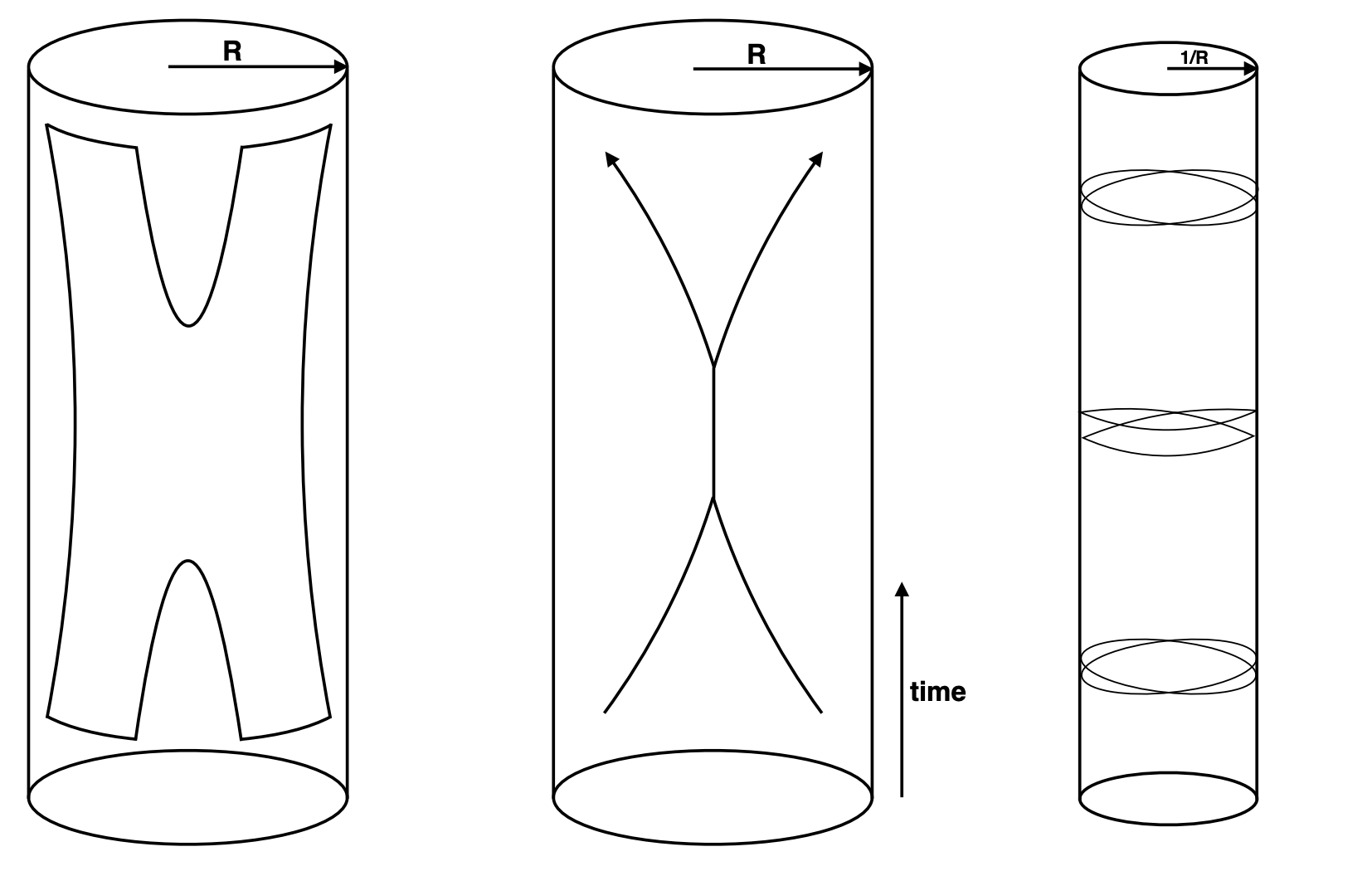}
  \centering
  \caption{}
  \label{T-duality}
\end{figure}

\begin{quote}
Scattering in a space with a closed dimension (other spatial dimensions
are not shown). In the center is the observed QFT process, in a dimension of
radius $R$, in which two particles interact to produce a third, which then decays
to two outgoing particles. On the left is the string description in a target space
of radius $R$: two strings join into one, which then splits into two. On the right
is the dual – physically equivalent – description in a target space of radius $1/R$:
two once-wound strings join to form a single twice-wound string, which then
splits back into two once-wound strings. \citep[p.~16]{HWforthcomingSTChapter9}
\end{quote}

\noindent In sum, there are alternative, equally good, ways to describe the same physical interactions. But they invoke different spacetimes, and different ontologies.

T-duality exists between type IIA and type IIB superstring theories, and between the two heterotic superstring theories ($SO\left(32\right)$ and $E_8 \times E_8$). For instance, type IIA superstring theory on a product manifold $M \times S^1$ with radius of the periodic dimension $R$, is dual to type IIB superstring theory on the product manifold $M \times S^1$ with radius of the periodic dimension proportional to $1/R$ \cite[ch.~6]{becker2006string}. 
   
%\subsection{Mirror Symmetry}

Mirror symmetry is a generalisation of T-duality to the case of \textit{topologically inequivalent target spaces}.\footnote{For a philosophical introduction to mirror symmetry, see \citet{rickles2013mirror}; \citet[section 3.2]{matsubara2018spacetime}.} A topological space is, intuitively, a structured set in which one can continuously deform subspaces. Any metric space is a topological space, but not all topological spaces are metric spaces. That dual target spaces may not only possess different compactification radii but even different topologies call even more into question that we could infer anything ontological about the physical world from the mathematical structure of the target spaces. Here is how \citet{matsubara2018spacetime} puts the point about inequivalent topologies, and the implications for our understanding of the effective spacetime (the standard old-fashioned spacetime we measure with rods and clocks):

\begin{quote}
%We now take a look at the more complicated dualities known as `mirror symmetries', in which different dual models can have target space manifolds with different topology. [\dots]
Spaces, or manifolds, with the same topology can be deformed into each other by stretching without cutting or tearing the manifold apart. Topology deals just with the overall way a space is connected, if it has holes in it for instance. A standard example is that a sphere cannot be deformed into a torus—the shape of a donut---without cutting or tearing. On the other hand the torus can be deformed into a coffee cup by stretching out a part from the torus and deforming it to the part of the cup that holds the coffee. The hole in the middle of the donut is preserved in the handle of the coffee cup.
Using the same kind of reasoning as in the T-duality case, what conclusions can be drawn regarding the effective spacetime when there is a duality of the mirror symmetry kind? \citep[p.~343]{matsubara2018spacetime}
\end{quote}

\noindent Thus, the question now is what kind of ontological implications one might find in situations where the target spaces are so radically different that they cannot be transformed into each other by smooth deformations.

Mirror symmetry is usually discussed by focusing on a particular class of those symmetries, involving Calabi-Yau spaces, but there exists other forms of mirror symmetry.\footnote{See \citet[p.~343]{matsubara2018spacetime} and references therein.} The study of Calabi-Yau spaces is motivated by the fact that they make it possible to compact six of the ten dimensions of space-time and explain why we seem to live in a four-dimensional world, while preserving supersymmetry. Being a generalisation of T-duality, mirror symmetry also exists between type IIA and type IIB superstring theories. As in the case of T-duality, if we consider a model (i.e. a solution) of type IIA string theory on a particular Calabi-Yau space, this solution will be fully physically equivalent to a solution of type IIB theory on a completely different, i.e. topologically non-equivalent, target space. \citeauthor{matsubara2018spacetime} nicely put the point as follows:

\begin{quote}
%The most studied examples of mirror symmetry deal with cases where six of the dimensions of the target spaces are compactified to form so-called `Calabi–Yau’. [\dots]
%In this discussion we will exclusively consider this form of mirror symmetry, but it should be noted that not all instances of mirror symmetry require that the compactified dimensions in the dual formulations are Calabi–Yau manifolds.[\dots] 
It has been found that dual, and thus physically equivalent, string theory models can be defined on pairs of target spaces which are the same with regards to their non-compactified four dimensions. However, their compactified dimensions form different Calabi–Yau manifolds $\mathcal{CY}_1$ and $\mathcal{CY}_2$ that are topologically inequivalent. Two such manifolds constitute a mirror pair. When the manifolds are exchanged, the type of string theory is also transformed so a type IIA string theory on $\mathcal{M}^4 \times \mathcal{CY}_1$ is equivalent to a type IIB string theory on $\mathcal{M}^4 \times \mathcal{CY}_2$; here $\mathcal{M}^4$ stands for the manifold describing the four non-compact dimensions. This is an even more drastic example compared to the T-duality case and it is even more difficult to make sense of what the spatiotemporal picture is supposed to be, because in this instance not even the topology is shared between the two dual target spaces. Furthermore, there is in contrast to the T-duality case no simple method of deriving what the relevant effective or phenomenal spatiotemporal picture would be.

Hence, if we can change certain topological properties of the compactified part of the target space manifold, while keeping the physics in the sense described above, the target spaces cannot be trusted to give even the right topology of spacetime due to the conflicting suggestions. \citep[p.~344]{matsubara2018spacetime}
\end{quote}

\noindent In short, T-duality shows that we should not take the geometry of target space at face value, i.e. as truly reflecting the physical facts about the geometry of our world (see \citealt{HuggettDualities} for a step-by-step demonstration). And mirror symmetry shows that this is not only true of the metric of the target space but even of its topology. We shall see in the next chapter what meaning we can give to this strange situation, and the arguments that can be mounted against the fundamentality of spacetime by drawing on T-duality and mirror symmetry.

\subsection{The AdS/CFT Correspondence}

Another duality involves not only string theories, but also quantum field theories---the type of theories involved in the Standard Model of particle physics. The central idea is that theories including a description of gravitation in $N+1$ dimensions are fully equivalent to non-gravitational quantum field theories in $N$ dimensions. The most important of these gauge/gravity correspondences is the \textit{AdS/CFT correspondence}.

On one side of the duality, we find a theory of quantum gravity: a type II superstring theory, defined on an AdS spacetime.\footnote{It can be a type IIA or type IIB theory, but also an M-theory or a low-energy effective description of type II string theory, called type II supergravity.} AdS stands for `Anti-de-Sitter' (an AdS spacetime is a vacuum solution to general relativity where the geometry has a constant negative curvature). Unfortunately, not only is our universe not empty, but cosmological data suggests that its curvature is positive. The geometry of our universe, neglecting its minute material content, should correspond approximately to a de Sitter (dS) spacetime, with constant positive curvature. Thus, one of the current challenges for string theorists is to move from the AdS/CFT correspondence to a dS/CFT correspondence or, alternatively, to show why the geometry of our real world is in fact AdS. This fact has led to some scepticism about the value of this duality, as it is difficult to see how the result could generalise to dS spacetimes, or why our real world would be AdS when it appears to be dS.%\footnote{There are various attempts to solve the difficulty, for instance by developing a \textit{Minkowski/CFT correspondence}; see \citet{de2003holographic}.}

On the other side of the duality is conformal field theory, that is, a particular type of field theory that has conformal symmetry---hence the acronyms `CFT'. Conformal symmetry means that the system is not sensitive to the size of the system in question, but only to the angles between the parts of the system. A conformal symmetry preserves angles, but not distances. A theory involving conformal symmetry is arguably strange, and conformal symmetry doesn't seem to be a property of any of the quantum field theories of the Standard Model of particle physics. But it's interesting to note that the CFT side can be used to perform important calculations. This is because the duality is a \textit{strong-weak} one. This means that a high-energy system on the CFT side corresponds to a low-energy system on the gravity side of duality, and vice-versa. Thus, if one wants to understand the high-energy behaviour of mass---involving gravity---one can use the CFT side, which does not involve gravity. Since high energy is difficult to describe---as perturbation theory, the standard approximation technique, cannot be used---strong-weak dualities are sometimes called hard-easy dualities. The AdS/CFT correspondence was found by \citet{maldacena1999large}, and regarded as implying a form of holography by \citet{witten1998anti}. A great deal of work is underway in physics to work out the details and to develop a number of more refined approaches.\footnote{A popular approach is \citet{ryu2006holographic}, who conjecture that the entanglement entropy in $d+1$-dimensional conformal field theories can be obtained from the area of d-dimensional surfaces in $AdS_{d+2}$. The claim is discussed in few philosophical works: \citet{jaksland2021entanglement, NeyPBS}. } %We will return to this question when we discuss the various ontological interpretations of duality in the next subsections.

The intuitive idea behind the AdS/CFT correspondence is that one can understand quantum gravity indirectly, in the gravitational bulk, by using a non-gravitational theory on the surface of the gravitational bulk. To understand this, think of a four-dimensional system like a star, and imagine it can be described as the four-dimensional boundary of a five-dimensional system, a star spread in five dimensions. The AdS/CFT correspondence generalizes this idea and implies that any physical system in five dimensions with gravitational properties can be described in four dimensions as the boundary of the five-dimensional bulk. 

As always, on these topics, our imagination is flawed, and we certainly can't produce mental images of five-dimensional stars. Thus, a little detour by a three-dimensional object might be useful. Earth is a three-dimensional object, and its surface is a two-dimensional surface. Usually, we think of this two-dimensional surface has having an extrinsic curve, namely a curve with respect to an extrinsic, embedding, three-dimensional space. But one could alternatively turn to Riemannian geometry and refuse the existence of any extrinsic embedding spacetime. The two-dimensional curvature of the Earth surface then becomes an intrinsic property, that one can record by measuring angles on its surface, without referring to any embedding, distinct geometry.

Imagine that we have two descriptions of Earth. One describes a bulk and includes everything that happens at its centre. The other is only about the surface of Earth and makes no reference to what is happening at its centre. In fact, there is no centre in this second description, because the surface is considered to have an intrinsic curvature only, so that there is no embedding space to make sense of an interior or exterior of the closed surface. If the AdS/CFT correspondence was applying to this case, it would be possible to describe everything that happens at the centre of the Earth, according to Theory 1, by simply describing what happens on its surface, as described by its dual Theory 2.

\begin{figure}[ht]
  \includegraphics[width=6cm]{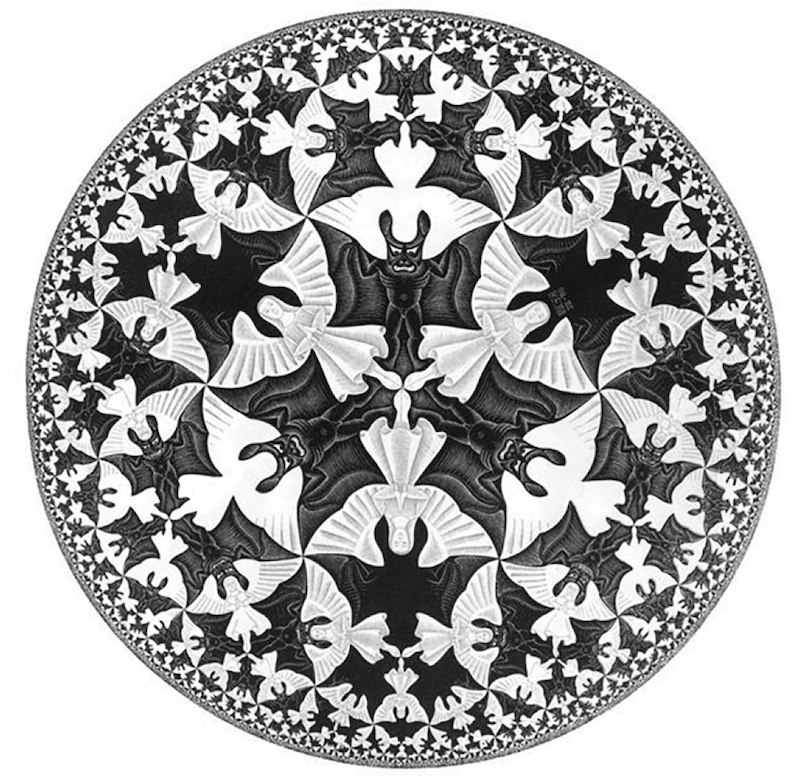}
  \centering
  \caption{The woodcut Circle limit IV by M. C. Escher. \textit{Pedro Ribeiro Simões / Flickr / CC BY 2.0}. }
  \label{adsangel}
\end{figure}

Figure \ref{adsangel} can be regarded as a perhaps even simpler illustration of the correspondence, by considering a two-dimensional $AdS_2$ spacetime, which has negative curvature by construction, and which is maximally filled of angels and demons. An $AdS_2/CFT_1$ correspondence would imply that all information about the angels and demons in the disk could be obtained from the properties of a world geometrically identical to its boundary surface.\footnote{For more details on the analogy, see \citet{dummies}, aka `String Theory for Dummies'.}

%In the classical case of Earth, this sounds crazy as, as a matter of fact, there is no relation strong enough to either move from the content of Earth to a description of the properties of its surface, or vice versa. But the situation is different in some contexts, and especially for black holes.

Now, a question one may perhaps ask is: how is it possible for a 10D superstring theory such as type IIB superstring theory to offer a 5D description dual to the 4D description of its boundary? The answer to this is that 10D spacetime can be analysed as a product space in different ways. A popular approach is to view the 10D spacetime as a $A d S_{5} \times \mathcal{S}^{5}$. The CFT thus describes a four-dimensional spacetime that corresponds to the boundary of the $A d S_{5}$ spacetime. The $\mathcal{S}^{5}$ structure on the bulk side then can be understood as an internal space (a second space, existing at each location in the bulk in the other space, and mathematically described by the bundle formalism) that corresponds to internal degrees of freedom on the CFT side, and the fifth dimension of the $A d S_{5}$ to an energy parameter on the CFT side. The boundary encodes both the extra radial dimension of $A d S_{5}$ and the compact extra dimensions $\mathcal{S}^{5}$ at each point of the bulk $A d S_{5}$. What the word `encodes' really means, in metaphysical terms, is a difficult question, and we will come back to it when discussing the metaphysics of duality in the next chapter.
%Duality between a theory that describes a world with gravity (an AdS spacetime) and a conformal field theory living on its lower dimensional boundary, a four-dimensional Minkowski spacetime without gravity.

%E.g. duality between IIB superstring theory and a Yang-Mills supersymmetric theory.

%The development of the AdS/CFT correspondence was essentially tied to black hole physics and black hole statistical mechanics \citep{wallaceBH}. Indeed, calculations made in the background of semi-classical gravity, usually viewed as a good approximation of low-energy quantum gravity, found that the number of micro-states of black holes seems is proportional to the event horizon surface rather than to the bulk volume of black holes. 

%This leads directly to two central ideas, one methodological, the other metaphysical. 

%Methodological

%The methodological idea is that one could use our understanding of the CFT side at low-energy 
%Starting point: Using conformal quantum field theory on the boundary of a higher dimensional spacetime described by string theory.

%Metaphysical

%\footnote{The AdS/CFT correspondence was originally introduced in \cite{Mal}.}
%This latter duality is popular in the context of discussion over black hole physics.
%\fnsep{In fact, all discussion in this paper applies to dualities \emph{simpliciter}, rather than specifically to string-theoretic dualities. Nevertheless, we focus upon the latter as a concrete case.}

\subsection{S-duality}

\textit{S-duality}, short for \textit{Sen duality} or \textit{strong-weak duality}, relates the solutions of one superstring theory with string coupling constant $g_s$ to the solutions of another superstring theory with string \textit{coupling constant} $1/g_s$ \citep{sen1994strong}. The coupling constant, or interaction constant, is the parameter in the equations that gives the strength of the interaction between the strings and fields. Basically, it describes the probability for each string to split or merge with other strings. It is thus a so-called `strong/weak' duality, just as the AdS/CFT correspondence (but the latter is not classified as a S-duality, even though both are cases of strong-weak dualities). Type I superstring theory is dual under S-duality to $SO(32)$ \cite[\S8.2]{becker2006string}. Type IIB is self-dual, which means that the duality map relates the solution space of type IIB to itself.

The fact that S-duality reverses the interaction constant between the two duals is a mathematical description. There is another easier, more ontological, way of describing S-duality. Remember that strings can attach to branes, and branes can be one-dimensional objects, just like strings. Thus, we can consider situations where open strings are attached to D1-branes. In this context, it is usual to refer to the open strings as `F-strings' (for \textit{fundamental} strings, the name has nothing to do with F-theory). Then, we can refer to composite objects as `($p$,$q$)-strings', where $p$ refers to the number of F-strings (fundamental strings) and $q$ to the number of D1-branes (or `D-strings') structuring those composite objects.

S-duality permutes $p$ and $q$ under the duality map, which raises doubts (once again!) about the idea of considering strings as more fundamental than branes. Similarly, there is no reason to consider branes as more fundamental than strings when S-duality is involved. Thus, as far as I can tell, S-duality offers an argument in favour of \textit{brane democracy} in string theory. As we will see in the next chapter, S-duality can also be used to question the mereological structure of particle physics, and the fact that some particles are elementary particles composing other composite particles.

%\subsection{More Than one Fundamental Theory?}

%In the next chapter, we will review several ontological interpretations of duality, as well as possible arguments based on these interpretations in support of the claim that spacetime is emergent in string theory.

\subsection{M-theory and F-theory}
\subsubsection{M-theory}

%\subsection{Supergravity (SUGRA)}

%Each of the five superstring theory can be associated an effective field theory (see chapter **REF). Basically, they all describe any system at low-energy. Interestingly, there is a sixth supergravity theory: 11D supergravity, conjectured to be the effective description of M-theory. The SUGRAs are very similar to general relativity, but with supersymmetry.

%\subsection{M-theory}

M-theory is usually regarded as more fundamental than the five superstring theories by being non-perturbative (i.e non-approximative), and more fundamental than eleven-dimensional supergravity, viewed as its low energy effective field description.\footnote{The theory has also been called `U-theory', for `Underlying' theory \citep[p.~37]{sen2001recent}.} %More concretely, it relates to the perturbative type IIA theory, via a form of dimensional reduction (M-theory and its eleventh dimension appears as a limit of type IIA when the string coupling goes to infinity). 
The theory was first envisaged as a potential candidate for an absolutely fundamental, final theory of everything. However, M-theory must factually be seen as the hypothesis of a theory that is relatively more fundamental than the theories of superstrings and supergravity. The fact that it should also be absolutely fundamental is a strong additional hypothesis that is not necessary, and perhaps is not that well motivated.
The `M' of M-theory has been variously described as referring to `magic', `matrix', `membrane' or `mother'. It does not describe anymore strings but higher-dimensional objects, the branes. Thus, if M-theory is on the right track, we should expect that strings are not (metaphysically) fundamental. 

The branes should be distinguished from the D-branes of the type I and II string theories, and for this purpose are called `M-branes'. They include two sorts of objects: M2-branes (two-dimensional surfaces aka membranes) and M5-branes (five-dimensional objects). The M2-branes can be regarded as roughly equivalent to the strings of the perturbative string theories, and the M5-branes as equivalent to their D-branes, by being five-dimensional sub-manifolds of the background space trapping the M2-branes. M-theory, strictly speaking, is\textit{ not} a string theory as it does not contain any elementary string in its ontology. However, the standard convention is to use the expression `string theory' to refer to the whole research programme, including the bosonic theories, the five superstring theories, 11D supergravity theories, the non-perturbative theories (M-theory and F-theory) and all the formalisms potentially relevant for the paradigm.

%Usually, dualities are introduced beforehand, to then motivate M-theory, thus following the historical order---as 
Witten conjectured the existence of M-theory in 1995 at the \textit{Strings Conference},  on the basis of the dualities (see e.g. \citealt{witten1995string}) the thought being that the existence of the duality web arises from the existence of the more fundamental M-theory. %However, I want to give a general presentation of string theory first, before going into the details of duality in the next section. %Indeed, duality will need to be presented in more detail to support the arguments that will be made in the next chapter concerning their ontological interpretations and their implications for the possible non-fundamentality of spacetime. 
%M-theory was first conjectured as a more fundamental, non-perturbative, theory by Witten to explain the many unexpected special relation that exist between the five superstring theories and supergravity. 

Here is a description of M-theory from \citet{zwiebach2009first}, a classical textbook:

\begin{quote}
    The limit of type IIA theory as the string coupling is taken to infinity was shown to give a theory in eleven dimensions. This theory is called M-theory, with the meaning of M to be decided when the nature of the theory becomes clear. It is known, however, that M-theory is \textit{not} a string theory. M-theory contains membranes (2-branes) and 5-branes, and these branes are not D-branes. M-theory may end up playing a prominent role in understanding string theory. The discovery of many other relationships between the five string theories listed above and M-theory has made it clear that we really have just \textit{one theory}. This is a fundamental result: there is a unique theory, and the five superstrings and M-theory are different limits of this unique theory. \citep[p.~325]{zwiebach2009first}
\end{quote}

\noindent We don't have any full understanding of M-theory but some of its models (i.e. solutions) can be constructed in different ways from the five superstring theories, by compactifying in different ways the eleventh dimension. On the one hand, M-theory with one dimension compactified on a circle gives rise to type IIA.\footnote{``M-theory compactified on a circle is type IIA superstring theory with some finite value for the string coupling'' \citep[p.~481]{zwiebach2009first}.} On the other hand, compactifying M-theory on a segment leads to the heterotic $E8 \times E8$ theory.\footnote{``Another popular approach that gives semi-realistic models uses a Calabi–Yau space to
reduce the spacetime dimensionality to five. The fifth dimension is then turned into a finite segment, a space that can be viewed as an orbifold (Problem 2.5). This is called the heterotic M-theory approach, because the compactification of M-theory on a segment has been shown to give the heterotic $E8 \times E8$ superstring'' \citep[p.~481]{zwiebach2009first}.} Also, 11-dimensional supergravity is considered a low-energy limit of M-theory. There exists a whole family of theories of supergravity, which are created by combining general relativity and supersymmetry. It was first thought that supergravity might offer a theory of everything. But now the prevailing view %since Witten's lecture 
is that 11-dimensional supergravity simply gives an approximate low-energy limit of M-theory \citep[p.~2]{blumenhagen2013basic}.

String theorists often claim that the five superstring theories are in fact one theory, as in the quote above. By this they mean that they offer insights into one theory, namely M-theory. However, another terminological usage, standard in philosophy of physics, is to consider them as separate theories bearing special relations (of duality) that will be discussed later. We will follow the standard philosophical terminology here and regard the different frameworks as different theories. 

Note also that in the last sentence of the previous quote, M-theory is explicitly presented not as being more fundamental than perturbative theories, but as another limit of a more fundamental theory. Hence the remark that the `M' in M-theory could mean several things. M-theory, could be a `mother theory' from which all the other perturbative theories derive, but that's not necessarily the case as the quote above shows. Yet, the majority view still seems to be that M-theory will turn out to be at least relatively more fundamental than the five perturbative theories. %---at least in the sense that it will be non-perturbative.

%BRANES

%Relations between M-theory and the other relations. Come back to this in section **REF other dualities, when discussing the relation between perturbative and non-perturbative theories. (should add at some point explanation on bosonic theories).

\subsubsection{F-theory}

Interestingly, there is another path to a non-perturbative theory, called `\textit{F-theory}' (see, e.g., \citet[section 18.8]{blumenhagen2013basic}; \citet{weigand2018tasi}; and \citet{CintiF-theory} for a philosophical presentation). The approach was introduced by \citet{vafa1996evidence}, when he established that certain solutions to type IIB superstring theory could be simplified in a new string theory with ten spatial dimensions and two temporal dimensions. F-theory is a non-perturbative approach to type IIB theory: 

\begin{quote}
Just as Special Relativity reduces to Newtonian mechanics in a specific asymptotic limit, notably the low-velocity limit ($v/c \rightarrow 0$), F-theory similarly reduces to Type IIB string theory within a certain asymptotic limit. Like M-theory, a proper non-perturbative definition of F-theory is not fully established. However, F-theory can be approached and examined from multiple perspectives, each elucidating different aspects and features of the theory. \citep[pp.~3-4]{CintiF-theory}
\end{quote}

It's also conjectured to be dual to M-theory.\footnote{See, e.g., \url{https://ncatlab.org/nlab/show/duality+between+M-theory+and+F-theory}.} One line of thinking for this possible duality is that type IIA and type IIB being T-dual, the duality could be uplifted from the perturbative to the non-perturbative level \citep[p.~5]{CintiF-theory}.

\begin{figure}[ht]
  \includegraphics[width=12cm]{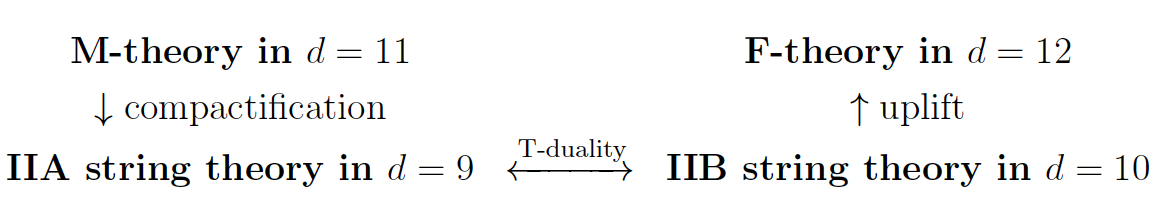}
  \centering
  \caption{\\ \small Is the duality at the perturbative level between Type IIA and Type IIB reflecting a duality between M-theory and F-theory? \textit{Credit: \citet[p.~5]{CintiF-theory}}}
  \label{web}
\end{figure}

\noindent The framework is sometimes regarded as a mathematical technique, rather than a theory. To quote Duff:

\begin{quote}
    The utility of \textit{F}-theory is beyond dispute and it has certainly enhanced our understanding of string dualities, Seiberg-Witten theory and much else. But should the twelve dimensions of F-theory be taken seriously? And if so, should F-theory be regarded as more fundamental than \textit{M}-theory? Given that there seems to be no supersymmetric field theory with \textit{SO}(10,2) Lorentz invariance \citep{nishino1996supersymmetric}, and given that the on-shell states carry only ten-dimensional momenta \citep{vafa1996evidence}, the more conservative interpretation is that the twelfth dimension is merely a mathematical artifact and that \textit{F}-theory should simply be interpreted as a clever way of compactifying the \textit{IIB} string \citep{sen1996f}. Time (or should I say `both times’?) will tell. \citep[pp.~327--328]{duff1999world}
\end{quote}

At any rate, F-theory is important as, first, it allows for non-perturbative calculations and, second---more interestingly for the philosopher---it potentially \textit{calls into question the very idea of the uniqueness of a more fundamental, non-perturbative, string theory}. If one takes seriously the existence of a non-perturbative F-theory, then it seems that there exists more than one non-perturbative string theory.

One way to reject that there is more than one fundamental non-perturbative theory would be to assert that the two theories are in fact aspects of an even more fundamental theory. Discussions about the relation between M-theory and F-theory are on-going, and some physicists believe that they are two aspects of the same theory \citep{FMtheory}. In philosophical parlance, this should be interpreted as the claim that M-theory and F-theory are relatively less fundamental than an even more fundamental theory. 

It's fair to say that to understand the nature of the relationship between M-theory and F-theory, we need to understand the nature of duality more generally. But as we shall see in the next chapter, the ontological status of duality is not clear because there exists a number of alternative ways of interpreting it. Many of these interpretations, we shall see, lead to arguments against the fundamentality of the entities laid down in dual descriptions, and in particular of spacetime.

\bibliography{references}
\bibliographystyle{chicago}
\end{document}